\documentclass[11pt]{article}

\usepackage{palatino}
\usepackage{mathpazo}
\usepackage{bbold}

\usepackage{amsfonts}
\usepackage{amssymb}
\usepackage{amsmath}
\usepackage{latexsym}
\usepackage{amsthm}
\usepackage{sectsty}

\newtheorem{theorem}{Theorem}
\newtheorem{lemma}[theorem]{Lemma}
\newtheorem{prop}[theorem]{Proposition}

\theoremstyle{definition}

\newcommand{\tinyspace}{\mspace{1mu}}

\newcommand{\op}[1]{\operatorname{#1}}
\newcommand{\norm}[1]{\left\lVert\tinyspace#1\tinyspace\right\rVert}
\newcommand{\abs}[1]{\left\lvert\tinyspace #1 \tinyspace\right\rvert}

\newcommand{\defeq}{\stackrel{\smash{\text{\tiny def}}}{=}}
\newcommand{\tr}{\operatorname{Tr}}
\renewcommand{\t}{{\scriptscriptstyle\mathsf{T}}}
\newcommand{\ip}[2]{\left\langle #1 , #2\right\rangle}

\newcommand{\setft}[1]{\mathrm{#1}}
\newcommand{\lin}[1]{\setft{L}\left(#1\right)}
\newcommand{\density}[1]{\setft{D}\left(#1\right)}

\newcommand{\herm}[1]{\setft{Herm}\left(#1\right)}
\newcommand{\pos}[1]{\setft{Pos}\left(#1\right)}

\newcommand{\ket}[1]{\vert #1 \rangle}
\newcommand{\bra}[1]{\langle #1\vert}
\def\I{\mathbb{1}}
\def\complex{\mathbb{C}}
\def\real{\mathbb{R}}
\def\natural{\mathbb{N}}

\newenvironment{mylist}[1]{\begin{list}{}{
	\setlength{\leftmargin}{#1}
	\setlength{\rightmargin}{0mm}
	\setlength{\labelsep}{2mm}
	\setlength{\labelwidth}{8mm}
	\setlength{\itemsep}{0mm}}}
	{\end{list}}

\newcommand{\class}[1]{\textup{#1}}

\def\X{\mathcal{X}}
\def\Y{\mathcal{Y}}
\def\A{\mathcal{A}}
\def\B{\mathcal{B}}

\begin{document}

  \title{\LARGE\bf
    Parallel approximation of non-interactive\\
    zero-sum quantum games}

  \author{
    Rahul Jain \quad\quad\quad John Watrous\\[1mm] 
    {\small\it Institute for Quantum Computing and School of Computer
      Science}\\ {\small\it University of Waterloo, Waterloo, Ontario,
      Canada}\\[1mm]
    {\small \{rjain,watrous\}@cs.uwaterloo.ca}
  }

  \date{August 20, 2008}

\maketitle

\begin{abstract}
  This paper studies a simple class of zero-sum games played by two
  competing quantum players: each player sends a mixed quantum state
  to a referee, who performs a joint measurement on the two states to
  determine the players' payoffs.
  We prove that an equilibrium point of any such game can be
  approximated by means of an efficient parallel algorithm, which
  implies that one-turn quantum refereed games, wherein the referee is
  specified by a quantum circuit, can be simulated in polynomial
  space.
\end{abstract}

\section{Introduction}

The theory of games has been studied extensively in mathematics and in
several other disciplines for which it has applications.
In theoretical computer science, computational aspects of game theory
represent an important focus of the field.

There are several settings of interest to quantum computation and
quantum cryptography that are naturally modeled by 
{\it quantum games}, which are games involving the exchange and
processing of quantum information.
For instance, multi-prover quantum interactive proofs
\cite{
  KobayashiM03,
  CleveHTW04,
  KempeKMTV08,
  KempeKMV08}
can be modeled as cooperative quantum games;
quantum coin-flipping 
\cite{
  Ambainis01,
  Kitaev02,
  SpekkensR02,
  Mochon04,
  Mochon07}
is naturally modeled as a game between two players that directly
exchange quantum information; 
and quantum refereed games
\cite{
  Gutoski05,
  GutoskiW05,
  GutoskiW07}
are competitive games that model quantum interactive proofs with
competing provers.

In this paper we consider a simple type of non-interactive, zero-sum
quantum game: two competing players (hereafter called {\it Alice} and
{\it Bob}) each send a mixed quantum state to a {\it referee}, who
performs a joint measurement on the two states to determine the
players' payoffs.
For a fixed description of the referee, let $\phi(\rho,\sigma)$ denote
Alice's expected payoff when she sends a mixed state $\rho$ to the
referee and Bob sends a mixed state $\sigma$.
(For zero-sum games, Bob's payoff is then given by $-\phi(\rho,\sigma)$.)
The theory of quantum information requires the function
$\phi(\rho,\sigma)$ to be bilinear, from which it follows that
\begin{equation} \label{eq:min-max-informal}
\max_{\rho} \min_{\sigma} \phi(\rho,\sigma) 
= \min_{\sigma} \max_{\rho} \phi(\rho,\sigma)
\end{equation}
from well-known variants of the Min-Max Theorem.
(Indeed, such a fact holds for a much more general class of quantum
zero-sum games that can allow for many rounds of interaction among the
referee and players \cite{GutoskiW07}.)
The value represented by the two sides of the equation
\eqref{eq:min-max-informal} is called the {\it value} of the game.
An {\it equilibrium point} of such a game is a pair of quantum states
$(\rho,\sigma)$ such that
\[
\max_{\rho'} \phi(\rho',\sigma) 
= \phi(\rho,\sigma) 
= \min_{\sigma'} \phi(\rho,\sigma'),
\]
the existence of which follows from the equation
\eqref{eq:min-max-informal}.
In other words, when one player plays one of the states of an
equilibrium point, the other has no incentive to play a state
different from the other state in the pair.
(These notions are, of course, similar to those for classical
zero-sum games, but with some technical differences due to the nature
of quantum information.
In particular, there is a continuum of pure strategies for quantum
players, corresponding to what are known as pure quantum states.)
An equilibrium point of a zero-sum quantum game, given as and explicit
description of the referee's measurement, can be efficiently computed
by means of semidefinite programming.

The main result of this paper is an efficient {\it parallel} algorithm
to find approximate equilibrium points of non-interactive zero-sum
quantum games.
For the case where the referee is specified by a quantum circuit
rather than in explicit matrix form, this algorithm implies that the
value of such a game can be approximated in polynomial space.
More succinctly, it implies that the complexity class $\class{QRG}(1)$
of problems having one-turn quantum refereed games is contained in
$\class{PSPACE}$.

Our algorithm is an example of the 
{\it multiplicative weights update method}, which is discussed in the
papers \cite{AroraHK05a,TsudaRW05}, for instance, and is explained in
detail in the PhD thesis of S.~Kale \cite{Kale07}.
This general method captures many previously discovered (and sometimes
re-discovered) algorithms, and has origins in learning theory, game
theory, and optimization.
The specific formulation of our algorithm is a non-commutative
extension of an (unpublished) algorithm of Rohit Khandekar and
the first author (Rahul Jain) that approximates equilibrium points of
classical games.

In the sections that follow, we give relevant definitions from the
theory of quantum information, present the algorithm and its analysis,
and discuss the containment $\class{QRG}(1)\subseteq\class{PSPACE}$
that follows.
We also explain how the problem of finding equilibrium points of
quantum games relates to the problem of approximating positive
instances of semidefinite programs.

\section{Preliminaries and definitions}\label{sec:definitions}

This section gives a brief summary of the quantum
information-theoretic concepts that are needed in the paper, and
then defines non-interactive zero-sum quantum games.
A few additional definitions that will be helpful later in the paper
are also discussed.

\subsection{Basic quantum information-theoretic notions}

In this paper we require just a few basic concepts about quantum
information; so it is not necessarily required that the reader has any
prior familiarity with it.

When we refer to a {\it quantum register} we simply mean a discrete
quantum system that we wish to consider, such as a collection of
qubits representing a message transmitted from one party to another.
With any quantum register we associate some vector space 
$\X = \complex^n$ for a positive integer $n$ that intuitively
represents the maximum number of distinct classical states that could
be stored in the register without error.
A {\it state} of such a register is represented by a 
{\it density matrix}, which is a $n\times n$ positive semidefinite
matrix having trace equal to 1.
Density matrices may reasonably be viewed as the quantum
information-theoretic analogue to a vector of probabilities,
representing a probability distribution.
We will write $\density{\X}$ to denote the set of all density matrices
associated with a register that is described by $\X$.
It is natural to view such density matrices as {\it linear operators}
acting on $\X$, and for this reason the term {\it density operator} is
commonly used in place of {\it density matrix}.

When two registers having associated spaces $\X = \complex^n$ and
$\Y=\complex^m$ are considered as a single compound register, the
associated space becomes the tensor product space
$\X\otimes\Y = \complex^{nm}$.
If the two registers are independently prepared in states described by
$\rho$ and $\sigma$, respectively, then the joint state is described by
the $nm\times nm$ density matrix $\rho\otimes\sigma$.
This matrix may be written in block form as
\[
\rho\otimes\sigma = \begin{pmatrix}
\rho_{1,1} \sigma & \cdots & \rho_{1,n} \sigma\\
\vdots & \ddots & \vdots \\
\rho_{n,1} \sigma & \cdots & \rho_{n,n} \sigma
\end{pmatrix}.
\]

In general, for a vector space $\X = \complex^n$, we write $\lin{\X}$
to denote the set of {\it all} $n\times n$ complex matrices, or linear
operators mapping $\X$ to itself, and we write 
$\herm{\X}$ to refer to the subset of $\lin{\X}$ given by the
{\it Hermitian} matrices.
These are the matrices $A$ satisfying $A = A^{\ast}$, where $A^{\ast}$
denotes the {\it adjoint} or {\it conjugate transpose} of $A$.
The set $\herm{\X}$ forms a vector space over $\real$, and many
optimization methods designed for real-valued symmetric matrices
extend to $\herm{\X}$ with little or no special consideration.
Finally, we write $\pos{\X}$ to denote the subset of $\herm{\X}$ that
consists of all  {\it positive semidefinite} $n\times n$ matrices (or
operators acting on $\X$).

The {\it Hilbert-Schmidt inner product} on $\lin{\X}$ is defined as
\[
\ip{A}{B} = \tr(A^{\ast} B)
\]
for all $A,B\in\lin{\X}$.
It holds that $\ip{A}{B}$ is a real number for all choices of
Hermitian matrices $A$ and $B$, and is a nonnegative real number for
all choices of positive semidefinite matrices $A$ and $B$.

A {\it measurement} of a register, having an associated vector space
$\X = \complex^n$, is described by a collection of $n\times n$
positive semidefinite matrices that sum to the identity.
Specifically, a measurement that has some finite, non-empty set
$\Sigma$ of possible outcomes is described by a collection
$\{P_a\::\:a\in\Sigma\}\subset\pos{\X}$ satisfying
\[
\sum_{a\in\Sigma} P_a = \I_{\X}.
\]
Here, $\I_{\X}$ denotes the $n\times n$ identity matrix, or identity
operator on $\X$.
(The subscript $\X$ is dropped when it is implicitly clear.)
If the register corresponding to $\X$ is in a state described by the
density matrix $\rho\in\density{\X}$, and this measurement described
by $\{P_a\::\:a\in\Sigma\}\subset\pos{\X}$ is performed, each outcome
$a\in\Sigma$ will be observed with probability $\ip{P_a}{\rho}$.

\subsection{Non-interactive zero-sum quantum games}

In a non-interactive zero-sum quantum game, Alice and Bob each send a
quantum state to a referee, who performs a measurement on these two
states to determine their payoffs.
Hereafter we will let $\A = \complex^n$ and $\B=\complex^m$
refer to the vector spaces corresponding to the states that Alice and
Bob send to the referee.

When the referee performs a measurement to determine Alice and
Bob's payoffs, a {\it joint} measurement is used.
In other words, Alice's and Bob's states are together viewed as a
single state of a register.
We therefore have that the referee's measurement is described by a
collection
\[
\{R_a\,:\,a\in\Sigma\} \subset \pos{\A\otimes\B}
\]
that satisfies the condition 
\[
\sum_{a\in\Sigma} R_{a} = \I_{\A\otimes\B}.
\]
If Alice sends the state $\rho$ and Bob sends the state $\sigma$, then each
possible measurement outcome $a\in\Sigma$ appears with probability
$\ip{R_a}{\rho\otimes\sigma}$.

A payoff for each player is associated with each possible
measurement outcome $a\in\Sigma$.
As we consider only zero-sum games, it is sufficient to describe these
payoffs by a function $v:\Sigma\rightarrow\real$;
with Alice's payoff for outcome $a$ being $v(a)$ and Bob's payoff
being $-v(a)$.
For a given choice of states $\rho$ and $\sigma$, it holds that Alice's
expected payoff is given by
\[
\sum_{a\in\Sigma} v(a) \ip{R_a}{\rho\otimes\sigma} 
= \ip{R}{\rho\otimes\sigma}
\]
for
\[
R = \sum_{a\in\Sigma} v(a) R_a.
\]
Bob's expected payoff is given by $-\ip{R}{\rho\otimes\sigma}$.
When one is interested only in the expected payoff of a given game, it
is therefore sufficient to consider that the game is simply determined
by $R$.
We will refer to $R$ as a {\it payoff observable}, given that a matrix
that arises in this way from a measurement and a real-valued function
on its outcomes is sometimes called an observable.

A necessary and sufficient condition for a matrix $R$ acting on
$\A\otimes\B$ to arise from some measurement and real-valued payoff
function $v$ as just described is that $R$ is Hermitian.
The sort of payoff function $\phi(\rho,\sigma)$ discussed in the
introduction therefore takes the form
$\phi(\rho,\sigma) = \ip{R}{\rho\otimes\sigma}$ for $R$ ranging over the set
of Hermitian matrices of the appropriate size.
As the tensor product is a universal bilinear function, and every
real-valued linear function on $\herm{\A}\otimes\herm{\B}$ can be
expressed as an inner product with some Hermitian matrix $R$, we have
that a necessary and sufficient condition for $\phi(\rho,\sigma)$ to be a
physically valid payoff function is that $\phi$ is a real-valued
bilinear function.

Now, given that the sets $\density{\A}$ and $\density{\B}$ are convex
and compact, and that Alice's expected payoff $\ip{R}{\rho\otimes\sigma}$
is a bilinear function on $\density{\A}\times\density{\B}$, it follows
from well-known extensions of von~Neumann's Min-Max Theorem
\cite{vonNeumann28} that
\begin{equation} \label{eq:min-max-R}
\max_{\rho\in\density{\A}} \min_{\sigma\in\density{\B}} 
\ip{R}{\rho\otimes\sigma}
=
\min_{\sigma\in\density{\B}} \max_{\rho\in\density{\A}} 
\ip{R}{\rho\otimes\sigma}.
\end{equation}
(See, for instance, \cite{Fan53}.)
We define $\alpha(R)$ to be the {\it value} of the game determined by
$R$, which is the quantity represented by the two sides of the above
equation \eqref{eq:min-max-R}.
A pair of quantum states $(\rho, \sigma)$ is called an 
{\it equilibrium point} for $R$ if both $\rho$ and $\sigma$
independently achieve the maximum and minimum, respectively, in
equation \eqref{eq:min-max-R}; or, equivalently, that
\[
\min_{\sigma' \in \density{\B}} \ip{R}{\rho\otimes\sigma'}
= \ip{R}{\rho\otimes\sigma} 
= \max_{\rho' \in \density{\A}} \ip{R}{\rho'\otimes\sigma}.
\]
Again, the existence of an equilibrium point follows easily from
equation \eqref{eq:min-max-R}.

We define that an {\it $\epsilon$-approximate equilibrium point} of a
game with payoff observable $R$ is a pair of states $(\rho,\sigma)$
such that
\[
\max_{\rho' \in \density{\A}} \ip{R}{\rho'\otimes\sigma} - \varepsilon
\norm{R}
\leq \ip{R}{\rho\otimes\sigma}
\leq \min_{\sigma' \in \density{\B}} \ip{R}{\rho\otimes\sigma'} +
\varepsilon \norm{R}.
\]
Note that this is an approximation in an additive sense, and is
relative to the maximum absolute value of any payoff (which is
reflected by the presence of the factor $\norm{R}$ in the error).

\subsection{Additional definitions and notation}
\label{sec:additional-notation}

This section summarizes some additional terminology and notation that
will be used in the paper.
First, a linear mapping of the form
$\Phi:\lin{\B}\rightarrow\lin{\A}$
is called a {\it super-operator} (as it maps linear operators to
linear operators).
The {\it adjoint super-operator} to $\Phi$ has the form
$\Phi^{\ast}: \lin{\A}\rightarrow\lin{\B}$,
and is uniquely determined by the condition
\[
\ip{A}{\Phi(B)} = \ip{\Phi^{\ast}(A)}{B}
\]
for all $A\in\lin{\A}$ and $B\in\lin{\B}$.
A super-operator $\Phi:\lin{\B}\rightarrow\lin{\A}$ is said to be
{\it positive} if it holds that $\Phi(P)$ is positive semidefinite for
every choice of a positive semidefinite operator $P\in\pos{\B}$.
It is the case that $\Phi^{\ast}$ is positive if and only if $\Phi$ is
positive.

There is a one-to-one and onto linear correspondence between
the collection of operators of the form $R\in\lin{\A\otimes\B}$ and
the collection of super-operators of the form
$\Phi:\lin{\B}\rightarrow\lin{\A}$, which is sometimes known as the
{\it Choi-Jamio{\l}kowski isomorphism}.
Specifically, for every super-operator
$\Phi:\lin{\B}\rightarrow\lin{\A}$, one defines an operator 
$R \in \lin{\A\otimes\B}$, called the Choi-Jamio{\l}kowski
representation of $\Phi$, by the equation
\[
R = \sum_{1\leq i,j \leq m} \Phi(E_{i,j}) \otimes E_{i,j}\;,
\]
where $E_{i,j}$ is the matrix with a 1 in entry $(i,j)$ and 0 in every
other entry.
Conversely, given an operator $R \in \lin{\A\otimes\B}$, one defines a
super-operator $\Phi:\lin{\B}\rightarrow\lin{\A}$ by means of the
formula
\begin{equation} \label{eq:CJ}
\Phi(B) = \tr_{\B} \left( R \left(\I_{\A}\otimes B^{\t}\right)\right).
\end{equation}
It follows that
\[
\tr\left(R(A\otimes B)\right) = \tr\left(A\,\Phi(B^{\t})\right)
\]
for every choice of $A\in\lin{\A}$ and $B\in\lin{\B}$.
These correspondences are both linear, and are inverse to one
another---so one is free to translate back and forth between the two
as necessary for a given application.
The assumption that $R$ is positive semidefinite implies that
the corresponding super-operator $\Phi$ is positive.
(In fact, $\Phi$ has the stronger property of being
{\it completely positive} if and only if $R$ is positive
semidefinite.)

For a given quantum game, we may equally well calculate expected
payoffs and equilibrium points by using the unique super-operator
$\Phi$ determined by \eqref{eq:CJ} rather than the payoff observable $R$.
In particular, $(\rho,\sigma^{\t})$ is an equilibrium point of the game
defining $R$ if and only if
\begin{equation} \label{eq:equilibrium-Phi}
\min_{\sigma' \in \density{\B}} \ip{\rho}{\Phi(\sigma')}
= \ip{\rho}{\Phi(\sigma)}
= \max_{\rho' \in \density{\A}}\ip{\rho'}{\Phi(\sigma)},
\end{equation}
and the value of this game is alternately expressed as
\begin{equation} \label{eq:value-Phi}
\alpha(\Phi) \defeq
\max_{\rho\in\density{\A}}
\min_{\sigma\in\density{\B}}
\ip{\rho}{\Phi(\sigma)} =
\min_{\sigma\in\density{\B}}
\max_{\rho\in\density{\A}}
\ip{\rho}{\Phi(\sigma)}.
\end{equation}

For a Hermitian $n\times n$ matrix $A$, one denotes the eigenvalues of
$A$ by
\[
\lambda_1(A) \geq \lambda_2(A) \geq \cdots \geq \lambda_n(A),
\]
sorted from largest to smallest and including each eigenvalue a number
of times equal to its multiplicity.
For every $n\times n$ Hermitian matrix $A$, the {\it spectral norm}
is denoted $\norm{A}$ and satisfies
\[
\norm{A} = \max\{\abs{\lambda_1(A)},\ldots,\abs{\lambda_n(A)}\},
\]
while the {\it trace norm} is denoted $\norm{A}_1$ and satisfies
\[
\norm{A}_1 = \abs{\lambda_1(A)}+ \cdots + \abs{\lambda_n(A)}.
\]
(Note that both of these formulas assume that $A$ is Hermitian.)

Using the above notation, we may express the equations
\eqref{eq:equilibrium-Phi} and \eqref{eq:value-Phi} in simpler terms:
$(\rho,\sigma^{\t})$ is an equilibrium point of the game defining $\Phi$
if and only if
\[
\lambda_1(\Phi(\sigma))
= \ip{\rho}{\Phi(\sigma)}
= \lambda_m(\Phi^{\ast}(\rho)),
\]
while the value of this game satisfies
\[
\alpha(\Phi) = 
\min_{\sigma\in\density{\B}}
\lambda_1(\Phi(\sigma))
=
\max_{\rho\in\density{\A}} 
\lambda_m(\Phi^{\ast}(\rho)).
\]

Finally, for future reference we note that if a payoff observable $R$
satisfies $0\leq R\leq \I$, then it holds that $0\leq \Phi(\sigma)\leq
\I$ and $0\leq \Phi^{\ast}(\rho)\leq \I$ for all choices of density
matrices $\rho\in\density{\A}$ and $\sigma\in\density{\B}$.
Moreover, for arbitrary Hermitian matrices $A\in\herm{\A}$ and
$B\in\herm{\B}$, we have $\norm{\Phi(B)} \leq \norm{B}_1$ and
$\norm{\Phi^{\ast}(A)} \leq \norm{A}_1$.

\section{The Main Result}
\label{sec:main}

We now present the main result of the paper, which is a parallel
algorithm to approximate the value of a non-interactive zero-sum
quantum game.
This fact is stated as Theorem~\ref{theorem:approximation}
below, following a few comments on the assumed form of the input.

We suppose that a given non-interactive zero-sum quantum game is
described by payoff observable $R\in\herm{\A\otimes\B}$, for 
$\A = \complex^n$ and $\B = \complex^m$ as discussed in the previous
section.
More precisely, we assume that $R$ is given as an $nm\times nm$
matrix, along with a specification of the dimensions $n$ and $m$.
Each entry of $R$ is a complex number, which we assume has rational
real and imaginary parts, each represented as the ratio of two
integers expressed in binary notation.
We let $k$ be the maximum length of the binary representation over all
of these integers, and define $\op{size}(R)$ to be $(nm)^2 k$.
It is clear that $O(\op{size}(R))$ bits suffice to encode~$R$.

In addition to $R$, $n$ and $m$, an accuracy parameter 
$\varepsilon > 0$ is also given as input.
For technical reasons it is most convenient to assume that
$\varepsilon$ is represented in {\it unary notation}: each string
$1^r$, for a positive integer $r$, denotes the value
$\varepsilon=1/r$.
This assumption on the input form of $\varepsilon$ reflects the fact
that our algorithm does not scale well with respect to accuracy---it
forces the length of the input to be proportional to $1/\varepsilon$
rather than $\log(1/\varepsilon)$, and therefore permits our algorithm
to be described by circuits with size polynomial in the input length.

The output of the algorithm will be a pair of density matrices
$(\rho,\sigma)$ where $\rho\in\density{\A}$ and $\sigma\in\density{\B}$.
They are assumed to be represented in a manner similar to the input
matrix $R$.

\begin{theorem}
\label{theorem:approximation}
An $\varepsilon$-approximate equilibrium point $(\rho,\sigma)$ for a 
given payoff observable $R$ can be computed 
by a logarithmic-space uniform family of Boolean circuits having depth
polynomial in $\log(\op{size}(R))$ and~$1/\varepsilon$.
\end{theorem}

\subsection{Parallel algorithm for positive games}
\label{sec:algorithm}

Our algorithm is most naturally described for the case that the payoff
observable $R$ satisfies $0 \leq R \leq \I$.
We therefore begin with this case, which will imply
Theorem~\ref{theorem:approximation} by an appropriate
translation and rescaling of $R$.
The algorithm is described in Figure~\ref{fig:algorithm}.

\begin{figure}[ht]
\noindent\hrulefill
\begin{center}
{\bf Algorithm}
\begin{enumerate}
\item
Let $\mu = \varepsilon/8$ and let
$N = \left\lceil 64 \ln(nm)/\varepsilon^2\right\rceil$.
\item
Initialize: $A_0 = \I_{\A}$, $\rho_0 = A_0/\tr(A_0)$,
$B_0 = \I_{\B}$, and $\sigma_0 = B_0/\tr(B_0)$.
\item
For each $j$ from 1 to $N$, let $A_j$, $\rho_j$, $B_j$, and $\sigma_j$
be as follows:
\begin{align*}
A_j & = \exp\left(\mu \sum_{i = 0}^{j-1} \Phi(\sigma_i)\right),\\
\rho_j & = A_j/\tr(A_j),\\
B_j & = \exp\left(-\mu \sum_{i = 0}^{j-1} \Phi^{\ast}(\rho_i)\right),\\
\sigma_j & = B_j/\tr(B_j).
\end{align*}

\item
Output the pair $\left(\rho,\sigma^{\t}\right)$, where
\[
\rho = \frac{1}{N} \sum_{j = 0}^{N-1} \rho_j 
\quad\quad \text{and} \quad\quad
\sigma = \frac{1}{N} \sum_{j = 0}^{N-1} \sigma_j.
\]
\end{enumerate}
\caption{A parallel algorithm for obtaining an
  $\varepsilon$-approximate equilibrium point of a one-round zero-sum
  quantum game.
  The game is assumed to be described by a payoff observable $R$
  satisfying $0\leq R\leq \I$, which gives rise to a positive map
  $\Phi:\lin{\B}\rightarrow\lin{\A}$ as described in
  Section~\ref{sec:additional-notation}.}
\label{fig:algorithm}
\end{center}
\noindent\hrulefill
\end{figure}

\subsubsection{Accuracy of the algorithm}

In this section, the accuracy of the algorithm described in
Figure~\ref{fig:algorithm} is analyzed.
We note that a similar type of analysis has appeared in previous works
on the multiplicative weights update method and its predecessors, and
in particular the reader is referred to \cite{Kale07} for information
on the generality of the approach.

At this point in the analysis we are concerned only with the idealized
algorithm described in Figure~\ref{fig:algorithm}---numerical issues
concerning the required precision with which the idealized operations
are performed are discussed in the next subsection.
We begin by noting some facts concerning matrix exponentials.
First, the {\it Golden-Thompson Inequality} 
(see Section IX.3 of \cite{Bhatia97}) states that, for any two
Hermitian matrices $X$ and $Y$ of equal dimension, we have
\[
\tr\left(e^{X + Y}\right) \leq \tr \left(e^X e^Y\right).
\]
Second is the following simple pair of inequalities concerning the
matrix exponential of positive and negative semidefinite matrices.

\begin{lemma} \label{lemma:exp-inequalities}
Let $P$ be an operator satisfying $0\leq P\leq \I$.  
Then for every real number $\mu > 0$, the following two inequalities
hold:
\begin{align*}
\exp(\mu P) & \leq \I + \mu \exp(\mu)P,\\
\exp(-\mu P) & \leq \I - \mu \exp(-\mu)P.
\end{align*}
\end{lemma}

\begin{proof}
It is sufficient to prove the inequalities for $P$ replaced by a
scalar $\lambda\in[0,1]$, for then the operator inequalities follow by
considering a spectral decomposition of $P$.
If $\lambda=0$ both inequalities are immediate, so let us assume
$\lambda>0$.
By the Mean Value Theorem there exists a value
$\lambda_0\in(0,\lambda)$ such that
\[
\frac{\exp(\mu \lambda) - 1}{\lambda} = \mu \exp(\mu \lambda_0)
\leq \mu \exp(\mu),
\]
from which the first inequality follows.
Similarly, there exists a value $\lambda_0\in(0,\lambda)$ such that
\[
\frac{\exp(-\mu\lambda)-1}{\lambda} = -\mu\exp(-\mu\lambda_0)
\leq -\mu \exp(-\mu),
\]
which yields the second inequality.
\end{proof}

We now proceed to the main part of the accuracy analysis,
which comprises two bounds on the eigenvalues of $\Phi^{\ast}(\rho)$
and $\Phi(\sigma)$, where $(\rho,\sigma^{\t})$ is the output of the algorithm.

\begin{lemma} \label{lemma:eigenvalue-bounds}
The following inequalities hold:
\begin{align}
\lambda_1(\Phi(\sigma)) & \leq \frac{\exp(\mu)}{N}\sum_{j = 1}^N 
\ip{\rho_{j-1}}{\Phi(\sigma_{j-1})} + \frac{\ln(n)}{\mu N},
\label{eq:A}\\
\lambda_m(\Phi^{\ast}(\rho)) & \geq \frac{\exp(-\mu)}{N}\sum_{j = 1}^N 
\ip{\rho_{j-1}}{\Phi(\sigma_{j-1})} - \frac{\ln(m)}{\mu N}.
\label{eq:B}
\end{align}
\end{lemma}

\begin{proof}
Let us begin by noting that each of the operators $A_j$ and $B_j$ 
(for $j = 0,\ldots,N$) that are obtained during the course of the
algorithm are positive definite, and therefore have positive trace.
It follows that each of the operators $\rho_j$ and $\sigma_j$ is a
well-defined density operator.

To prove the first inequality, observe that
\[
A_N = \exp\left(\mu\sum_{j = 1}^N \Phi(\sigma_{j-1})\right)
= \exp\left(\mu N \Phi(\sigma)\right).
\]
Given that $A_N$ is positive definite, it holds that
\[
\tr(A_N) \geq \lambda_1(A_n) = \exp(\mu N\lambda_1(\Phi(\sigma))),
\]
and therefore
\begin{equation} \label{eq:X-inequality-1}
\lambda_1(\Phi(\sigma)) \leq \frac{\ln(\tr(A_N))}{\mu N}.
\end{equation}

The inequality \eqref{eq:A} will now follow by bounding
$\ln(\tr(X_N))$, which can be done as follows.
First, note that we may alternately write
\[
A_j = \op{exp}\left(\ln(A_{j-1}) + \mu\,\Phi(\sigma_{j-1})\right)
\]
for each $j\geq 1$, and therefore
\[
\tr(A_j) = \tr\left(\op{exp}\left(\ln(A_{j-1}) 
+ \mu\,\Phi(\sigma_{j-1})\right)\right)
\leq \tr\left(A_{j-1} 
\exp\left(\mu\,\Phi(\sigma_{j-1})\right)\right)
\]
by the Golden-Thompson inequality.
As $\sigma_{j-1}$ is a density operator, it holds that
$\Phi(\sigma_{j-1})\leq \I$, and therefore
\[
\exp\left(\mu\,\Phi(\sigma_{j-1})\right)
\leq \I + \mu\exp(\mu)\Phi(\sigma_{j-1})
\]
by Lemma~\ref{lemma:exp-inequalities}.
Thus, using the fact that $\tr(X Y_1)\leq \tr(X Y_2)$ for all choices
of matrices $X$, $Y_1$, and $Y_2$ with $X\geq 0$ and $Y_1\leq Y_2$, we
have
\[
\tr(A_j)
\leq
\tr\left(A_{j-1}\left(\I+\mu\exp(\mu)\Phi(\sigma_{j-1})\right)\right)
=
\tr(A_{j-1})\left(1+\mu\exp(\mu)\ip{\rho_{j-1}}{\Phi(\sigma_{j-1})}\right).
\]
It now follows from the inequality $1 + \mu \leq \exp(\mu)$ that
\[
\tr(A_j)
\leq \tr(A_{j-1})\exp\left(\mu \exp(\mu) 
\ip{\rho_{j-1}}{\Phi(\sigma_{j-1})}\right).
\]
Applying this inequality recursively, and using the fact that
$\tr(A_0) = n$, we obtain
\begin{equation} \label{eq:X-inequality-2}
\tr(A_N) \leq 
\exp\left(\mu\exp(\mu)
\sum_{j = 1}^N \ip{\rho_{j-1}}{\Phi(\sigma_{j-1})} + \ln(n)\right).
\end{equation}
Combining \eqref{eq:X-inequality-1} and \eqref{eq:X-inequality-2}
yields
\[
\lambda_1(\Phi(\sigma)) \leq
\frac{\exp(\mu)}{N}\sum_{j = 1}^N 
\ip{\rho_{j-1}}{\Phi(\sigma_{j-1})} + \frac{\ln(n)}{\mu N},
\]
as required.

The second inequality follows by similar reasoning, except with a few
differences that we now highlight.
We first observe that
\[
B_N = \exp\left(-\mu N \Phi^{\ast}(\rho)\right).
\]
This time we have
\[
\tr(B_N) \geq \lambda_1(B_N) = \exp(-\mu N \lambda_m(\Phi^{\ast}(\rho))),
\]
where the switch from the largest eigenvalue to the smallest is
caused by the minus sign in the exponential function.
Thus,
\begin{equation}\label{eq:Y-inequality-1}
\lambda_m(\Phi^{\ast}(\rho)) \geq -\frac{\ln(\tr(B_N))}{\mu N}.
\end{equation}
The quantity $\tr(B_N)$ is now bounded in the same way as $\tr(A_N)$,
except that we need the second inequality in
Lemma~\ref{lemma:exp-inequalities}.
Specifically, the Golden-Thompson inequality implies
\[
\tr(B_j) = \tr(\exp(\ln(B_{j-1}) - \mu\Phi^{\ast}(\rho_{j-1})))
\leq \tr(B_{j-1}\exp(-\mu\Phi^{\ast}(\rho_{j-1}))).
\]
As $\Phi^{\ast}(\rho_{j-1})\leq\I$ we have
\[
\exp(-\mu\Phi^{\ast}(\rho_{j-1})) 
\leq \I - \mu\exp(-\mu) \Phi^{\ast}(\rho_{j-1}),
\]
and therefore
\[
\tr(B_j) \leq \tr(B_{j-1})\exp(-\mu\exp(-\mu)
\ip{\sigma_{j-1}}{\Phi^{\ast}(\rho_{j-1})}).
\]
It follows that
\begin{equation}\label{eq:Y-inequality-2}
\tr(B_N)\leq \exp\left(-\mu\exp(-\mu)
\sum_{j=1}^N \ip{\sigma_{j-1}}{\Phi^{\ast}(\rho_{j-1})} + \ln(m)\right).
\end{equation}
Combining \eqref{eq:Y-inequality-1} and \eqref{eq:Y-inequality-2},
along with the fact that 
$\ip{\sigma_j}{\Phi^{\ast}(\rho_j)} = \ip{\rho_j}{\Phi(\sigma_j)}$ for every
choice of $j$, yields
\[
\lambda_m(\Phi^{\ast}(\rho)) 
\geq \frac{\exp(-\mu)}{N}\sum_{j = 1}^N 
\ip{\rho_{j-1}}{\Phi(\sigma_{j-1})} - \frac{\ln(m)}{\mu N}
\]
and completes the proof.
\end{proof}

It is now possible to verify that the output $(\rho,\sigma^{\t})$ of
the algorithm satisfies
\[
\max_{\rho' \in \density{\A}} \ip{R}{\rho'\otimes\sigma^{\t}} - \varepsilon/2
\leq \ip{R}{\rho\otimes\sigma^{\t}}
\leq \min_{\sigma' \in \density{\B}} \ip{R}{\rho\otimes\sigma'}+\varepsilon/2,
\]
which is expressed in terms of the mapping $\Phi$ as
\[
\lambda_1(\Phi(\sigma)) - \varepsilon/2
\leq  \ip{\rho}{\Phi(\sigma)} 
\leq \lambda_m(\Phi^{\ast}(\rho)) + \varepsilon/2.
\]
It follow from Lemma~\ref{lemma:eigenvalue-bounds} that
\[
\lambda_1(\Phi(\sigma)) - \lambda_m(\Phi^{\ast}(\rho)) \leq
\frac{\exp(\mu) - \exp(-\mu)}{N}
\sum_{j = 1}^N\ip{\sigma_{j-1}}{\Phi(\rho_{j-1})} +
\frac{\ln(nm)}{\mu N},
\]
and given that each of the quantities
$\ip{\sigma_{j-1}}{\Phi(\rho_{j-1})}$ is at most 1, we have
\[
\lambda_1(\Phi(\sigma)) - \lambda_m(\Phi^{\ast}(\rho)) 
\leq 2\sinh(\mu) + \frac{\ln(nm)}{\mu N}
< 3\mu + \frac{\ln(nm)}{\mu N} \leq \varepsilon/2.
\]
Thus, given that 
$\lambda_m(\Phi^{\ast}(\rho)) \leq \ip{\rho}{\Phi(\sigma)} \leq
\lambda_1(\Phi(\sigma))$, we have
\begin{equation} \label{eq:final-accuracy}
\lambda_1(\Phi(\sigma)) - \varepsilon/2 
\leq \lambda_m(\Phi^{\ast}(\rho))
\leq \ip{\rho}{\Phi(\sigma)} 
\leq \lambda_1(\Phi^{\ast}(\rho)) 
\leq \lambda_m(\Phi(\sigma)) + \varepsilon/2
\end{equation}
as claimed.

\subsubsection{Numerical precision and complexity of the algorithm}

Let us now consider the complexity of the algorithm described in
Figure~\ref{fig:algorithm}.
It is the goal of this section to demonstrate that this algorithm can
be implemented, by a logarithmic-space uniform family of Boolean
circuits with depth polynomial in $\log(\op{size}(R))+1/\varepsilon$,
with sufficient accuracy to obtain an $\varepsilon$-approximate
equilibrium point for the input payoff observable $R$.
Throughout the analysis, we (sometimes grossly) overestimate errors
for the sake of simpler expressions involving as few variables as
possible.

Each iteration performed in step 3 of the algorithm requires the
evaluation of $\Phi$ and $\Phi^{\ast}$, two matrix exponential
computations, and a constant number of elementary matrix operations
(in this case: addition, scalar multiplication, and computation of the
trace).
Were it not for the matrix exponentials, it would be straightforward
to perform all of the required operations within the claimed size and
depth bounds using exact computations.
Given that the matrix exponentials will generate irrational numbers,
however, we must settle for approximations over the course of the
algorithm.
To guarantee that the algorithm is sufficiently accurate, it will
suffice to perform all computations to within an additive error of
$(\varepsilon/2)\exp(-8N^2)$, as is shown below.
(We could afford to take a much smaller error with respect to
$\op{size}(R)$, but there is no need to do this.)

Let us begin by making a few simple observations about the matrices
computed throughout the course of the algorithm.
The matrices $\sigma_0,\ldots,\sigma_{N-1}$ are density matrices, and
therefore it holds that $\norm{\Phi(\sigma_j)}\leq 1$ for each choice of
$j = 0,\ldots,N-1$.
Likewise, $\norm{\Phi^{\ast}(\rho_j)}\leq 1$ for each choice of 
$j = 0,\ldots,N-1$.
Consequently, we have $\norm{A_j}\leq e^{N}$,
$\norm{B_j}\leq e^{N}$,
\[
1\leq \tr(A_j)\leq e^{2N}\quad\quad\text{and}\quad\quad
e^{-N} \leq \tr(B_j) \leq e^{N}
\]
for $j = 0,\ldots,N-1$.

Next, let us represent the actual matrices computed during the course of the
algorithm by placing a tilde over the variables representing the
idealized values that are expressed in Figure~\ref{fig:algorithm}.
It will suffice to prove that
$\norm{\rho - \widetilde{\rho}}_1 \leq \varepsilon/2$ and 
$\norm{\sigma - \widetilde{\sigma}}_1 \leq \varepsilon/2$, for then the
inequalities
\[
\abs{\lambda_1(\Phi(\sigma)) -
  \lambda_1\left(\Phi\left(\widetilde{\sigma}\right)\right)} \leq
\frac{\varepsilon}{2}
\quad\quad\text{and}\quad\quad
\abs{\lambda_m(\Phi^{\ast}(\rho)) -
  \lambda_m\left(\Phi^{\ast}\left(\widetilde{\rho}\right)\right)} 
\leq \frac{\varepsilon}{2}
\]
hold.
Combined with \eqref{eq:final-accuracy}, we obtain the required
accuracy.

Now, each iteration of step 3 of the algorithm will introduce some
error into the calculation of the final answer.
Let us consider the $j$-th iteration, and assume that a positive real
number $\delta_j\in(0,1)$ is given such that
$\norm{\sigma_i - \widetilde{\sigma}_i}_1 \leq \delta_j$ for 
$i = 0,\ldots,j-1$.
Let us define
\[
X_j = \mu \sum_{i = 0}^{j-1}\Phi(\sigma_i)
\quad\quad\text{and}\quad\quad
\widetilde{X}_j 
= \mu \sum_{i = 0}^{j-1}\Phi\left(\widetilde{\sigma}_i\right).
\]
Then $\norm{X_j - \widetilde{X}_j} \leq \delta_j N$ and $\norm{X_j}\leq
N$, and therefore
\[
\norm{\exp(X_j) - \exp\left(\widetilde{X}_j\right)} \leq 
\norm{X_j - \widetilde{X}_j} e^{\norm{X_j-\widetilde{X}_j}} e^{\norm{X_j}}
< \delta_j e^{3N},
\]
where the first inequality follows from Corollary 6.2.32 of
\cite{HornJ91}.
By computing the matrix exponential with accuracy $\delta_j$ we
therefore have $\norm{A_j - \widetilde{A}_j} \leq \delta_j e^{4N}$,
and thus $\norm{A_j - \widetilde{A}_j}_1 \leq \delta_j e^{5N}$.
It follows that
\[
\norm{\rho_j - \widetilde{\rho}_j}_1
\leq
\frac{1}{\tr(A_j)} \norm{A_j - \widetilde{A}_j}_1
+ \norm{\widetilde{A}_j}_1 \abs{\frac{1}{\tr(A_j)} -
  \frac{1}{\tr\left(\widetilde{A}_j\right)}} \leq \delta_j e^{8N}.
\]
By similar reasoning, if it holds that
$\norm{\rho_i - \widetilde{\rho}_i}_1 \leq \delta_j$ for 
$i = 0,\ldots,j-1$, then
\[
\norm{\sigma_j - \widetilde{\sigma}_j}_1
\leq
\frac{1}{\tr(B_j)} \norm{B_j - \widetilde{B}_j}_1
+ \norm{\widetilde{B}_j}_1 \abs{\frac{1}{\tr(B_j)} -
  \frac{1}{\tr\left(\widetilde{B}_j\right)}} \leq \delta_j e^{8N}.
\]
We conclude from these bounds that taking $\delta_j =
(\varepsilon/2)e^{-8 N^2}$ guarantees that
$\norm{\rho - \widetilde{\rho}}_1 \leq \varepsilon/2$ and 
$\norm{\sigma - \widetilde{\sigma}}_1 \leq \varepsilon/2$.

The required precision for the matrix exponentials is easily obtained
by taking sufficiently many terms in the series
$e^X = \I + X + X^2/2 + X^3/6 + \cdots$.
(This of course is not the most efficient way to compute matrix
exponentials, but it suffices to prove the main theorem.)
For instance, taking $9N^2$ terms guarantees that the required
accuracy $(\varepsilon/2)e^{-8N^2}$ is achieved.

At this point, the parallel complexity of the algorithm is easily
bounded.
Each of the matrices stored by the algorithm has entries whose
real and imaginary parts are represented in binary notation using
$O(N^2)$ bits.
For each iteration in step 3 of the algorithm, the evaluations of
$\Phi$ and $\Phi^{\ast}$, as well as the elementary matrix
operations, may therefore be performed by standard parallel algorithms
(see, for instance, \cite{vzGathen93}) by logarithmic-space uniform
Boolean circuits (with size that is necessarily polynomial in
$\op{size}(R)$ and $1/\varepsilon$ given this uniformity constraint),
within depth that is polynomial in $\log(\op{size}(R))$ and
$1/\varepsilon$.
The number of iterations performed is $N$, which results in total
depth polynomial in $\log(\op{size}(R))$ and $1/\varepsilon$.

\subsection{Extensions to arbitrary payoff observables}

For an arbitrary payoff observable $R$, the algorithm from the
previous section is not guaranteed to function correctly, as we have
used the positivity of the corresponding super-operator $\Phi$ several
times during the analysis.

It is straightforward, however, to translate and scale an arbitrary
payoff observable in a way that allows the algorithm to be used.
For an arbitrary positive semidefinite payoff observable $R$, this is
essentially trivial---one simply runs the algorithm on the payoff
observable $P = R/\norm{R}$.
For a negative semidefinite payoff observable $R$, one simply exchanges
the roles of Alice and Bob and considers the payoff observable $-R$
(with the spaces $\A$ and $\B$ swapped).

Let us now consider the general case of a payoff observable $R$ for
which $\lambda_1(R) > 0 > \lambda_{nm}(R)$.
Define
\[
P  = \frac{R - \lambda_{nm}(R)\I}{\lambda_1(R) - \lambda_{nm}(R)}.
\]
Then $0\leq P \leq \I$, and so the algorithm from the previous section
may be used to obtain an $\varepsilon$-approximation $(\rho,\sigma)$
for $P$.
The point $(\rho,\sigma)$ is easily verified to be a $\delta$-approximate
equilibrium point for $R$, where
\[
\delta = \frac{\lambda_1(R) - \lambda_{nm}(R)}{\norm{R}}\,\varepsilon
\leq 2 \varepsilon.
\]

\section{$\class{QRG(1)}$ is contained in \class{PSPACE}} 
\label{sec:qsp2}

Quantum interactive proof systems with two competing provers are
naturally represented as games between two competing players,
moderated by a referee.
The two players (Alice and Bob) play the roles of competing provers,
while the referee corresponds to the verifier.
Quantum refereed games have been studied in
\cite{GutoskiW05,Gutoski05,GutoskiW07}, and represent a quantum
analogue to the classical refereed games model studied in
\cite{FeigeK97}.

The simplest form of a refereed quantum game has the general form
defined in Section~\ref{sec:definitions}; meaning that there is no
communication from the referee to the players.
The players each send a quantum state and the referee measures to
determine the winner.
With this picture in mind, one defines the complexity class
$\class{QRG(1)}$ to be the class consisting of all promise problems
$A = (A_{\mathrm{yes}},A_{\mathrm{no}})$ for which there exists a
polynomial-time uniform family $Q = \{Q_n\,:\,n\in\natural\}$ of
quantum circuits, where each circuit $Q_n$ takes $n + 2 p(n)$ input
qubits for some polynomial bounded function $p$, such that the
following properties hold:
\begin{mylist}{\parindent}
\item[$\bullet$] For every string $x\in A_{\mathrm{yes}}$ it holds
  that
  \[
  \max_\rho \min_\sigma \op{Pr}[Q(x,\rho,\sigma) = 1]  
  \geq \frac{2}{3}.
  \]
\item[$\bullet$] For every string $x\in A_{\mathrm{no}}$ it holds
  that
  \[
  \max_\rho \min_\sigma \op{Pr}[Q(x,\rho,\sigma) = 1]  
  \leq \frac{1}{3}.
  \]
\end{mylist}
Here, the maximum and minimum are both over all quantum states on
$p(\abs{x})$ qubits, and the notation $Q(x,\rho,\sigma) = 1$ is shorthand
for the event that a measurement of some fixed output qubit of the
circuit $Q_{\abs{x}}$ (with respect to the standard basis) yields 1,
assuming that the input to the circuit is the state
$\ket{x}\bra{x}\otimes \rho\otimes \sigma$.
The name $\class{QRG}(1)$ refers to the fact that these are quantum
refereed games with 1 turn, during which the players send quantum
states to the referee in parallel.

The class $\class{QRG}(1)$ may be viewed as a simple variant of
$\class{QMA}$, where there are two competing provers rather than a
single prover.
It is obvious that $\class{QMA}\subseteq\class{QRG(1)}$, and that
$\class{QRG}(1)$ is closed under complementation (and thus
$\class{co-QMA}\subseteq\class{QRG}(1)$).

The class $\class{QRG}(1)$ may roughly be thought of as a quantum
analogue to the class $S_2^P$ that was defined by \cite{Canetti96} and
\cite{RussellS98}, and it is easily observed that
$S_2^P\subseteq\class{QRG}(1)$.
There is one subtlety, however, which is that the definition of
$\class{QRG}(1)$ does not allow one prover to see the other's message
(which would not make sense in the quantum setting anyway), whereas 
the standard definition of $S_2^P$ does.

\begin{prop}
$\class{QRG}(1) \subseteq\class{PSPACE}$.
\end{prop}

\begin{proof}
Suppose that $A$ is a promise problem contained in $\class{QRG}(1)$,
and that $\{Q_n\}$ is a polynomial-time uniform family of quantum
circuits that witnesses this fact.
For each input $x$, let $R_x$ denote the payoff observable that
corresponds to the game played by the players Alice and Bob on input
$x$, where the payoff for Alice is defined as 1 for acceptance and 0
for rejection.
The expected payoff is therefore the probability of acceptance, which
Alice tries to maximize and Bob tries to minimize.

We denote by $\class{NC}(\mathit{poly})$ the class
of promise problems computed by polynomial-space uniform Boolean
circuits with polynomial depth.
It holds that $\class{NC}(\mathit{poly}) \subseteq \class{PSPACE}$
\cite{Borodin77}, so it therefore suffices to prove that
$A\in\class{NC}(\mathit{poly})$.
This is easily accomplished by composing three families of Boolean
circuits:
\begin{mylist}{\parindent}
\item[1.]
A family of Boolean circuits that outputs a description of
the payoff observable $R_x$ associated with the game on input $x$.
\item[2.]
The family of Boolean circuits given by
Theorem~\ref{theorem:approximation}, that finds an
$\varepsilon$-approximate equilibrium point $(\rho,\sigma)$ of the payoff
observable $R_x$, for $\varepsilon = 1/8$.
\item[3.]
A family of Boolean circuits that computes the expected payoff
$\ip{R_x}{\rho\otimes\sigma}$, and accepts if the value is greater than
1/2 (rejecting otherwise).
\end{mylist}

The first family is easily derived from the circuits $\{Q_n\}$,
by computing the product of a polynomial number of exponential-size
matrices that correspond to the quantum gates of the appropriate
circuit $Q_n$.
This family may be taken to be polynomial-space uniform, with
polynomial depth.
The second family is, as suggested above, given by
Theorem~\ref{theorem:approximation}.
This family is logarithmic-space uniform and has polynomial-size and
poly-logarithmic depth with respect to $\op{size}(R_x)$.
Thus, with respect to the input length $\abs{x}$, this family is
polynomial-space uniform, and has polynomial depth.
The last family is required only to perform elementary matrix and
arithmetic operations, and can be taken to have similar properties as
the first two: polynomial-space uniformity and polynomial depth.
Composing these families appropriately demonstrates that
$A\in\class{NC}(\mathit{poly})$ as required.
\end{proof}

\section{Parallel Approximation of Positive Semidefinite Programs} 
\label{sec:positive-SDPS}

We now discuss the connection between equilibrium points of
non-interactive zero-sum quantum games and semidefinite programs.
The main focus of this section will, in particular, be on
{\it positive} instances of semidefinite programs, and on the question
of whether good parallel methods to approximate them exist.
We will first discuss the general notion of positive instances of
semidefinite programs and then explain how our algorithm may be used
in their approximation, albeit with poor accuracy in some cases.

The multiplicative weights update method has been applied to
semidefinite programming in \cite{AroraHK05b,AroraK07}, and the
general connection between equilibrium points of different types
of games and linear/semidefinite programs is well-known.
Once again, the reader is referred to Kale \cite{Kale07} for further
details and historical remarks.

\subsection{Positive instances of semidefinite programs in super-operator form}

Suppose that the following input has been given:
\begin{enumerate}
\item
  a Hermitian matrix $A\in\complex^{n\times n}$,
\item
  a Hermitian matrix $B\in\complex^{m\times m}$, and
\item
  a linear mapping 
  $\Phi:\complex^{m\times m}\rightarrow\complex^{n\times n}$ (i.e., a
  super-operator) that preserves Hermiticity.
\end{enumerate}
To say that $\Phi$ preserves Hermiticity means that $\Phi(Y)$ is
Hermitian for every choice of a Hermitian matrix
$Y\in\complex^{m\times m}$.
This condition is equivalent to the Choi-Jamio{\l}kowski
representation $R$ of $\Phi$ being a Hermitian matrix.

Given this input, let us consider the following semidefinite
programming problem, which we say is in the {\it super-operator form}:
\begin{center}
  \begin{minipage}{3in}
    \centerline{\underline{Super-operator primal form}}\vspace{-7mm}
    \begin{align*}
      \text{maximize:}\quad & \ip{B}{Y}\\
      \text{subject to:}\quad & \Phi(Y) \leq A,\\
      & Y\geq 0.
    \end{align*}
  \end{minipage}
  \begin{minipage}{3in}
    \centerline{\underline{Super-operator dual form}}\vspace{-7mm}
    \begin{align*}
      \text{minimize:}\quad & \ip{A}{X}\\
      \text{subject to:}\quad & \Phi^{\ast}(X) \geq B,\\
      & X\geq 0.
    \end{align*}
  \end{minipage}
\end{center}
Here, $X$ and $Y$ range over all (positive semidefinite) matrices in
$\complex^{n\times n}$ and $\complex^{m\times m}$, respectively.
This form is completely general: it is possible to translate
semidefinite programs in so-called {\it standard form} to the
super-operator form, and vice versa.
It can be shown that strong duality holds for semidefinite programs in
the super-operator form under conditions that are similar to those for
standard form semidefinite programs.
In particular, the existence of either of the following implies that
strong duality holds:
\begin{enumerate}
\item
a positive definite matrix $Y$ for which $\Phi(Y) < A$, or
\item
a positive definite matrix $X$ for which $\Phi^{\ast}(X) > B$.
\end{enumerate}

We define that such a problem instance is {\it positive} if $A$ and
$B$ are positive semidefinite matrices and $\Phi$ is a positive
super-operator.
Let us also define that such a problem is {\it strictly positive} if it
holds that $A$ and $B$ are positive definite and $\Phi$ is a strictly
positive super-operator (which means that $\Phi(\I)$ is positive
definite).
Strong duality necessarily holds for all strictly positive
semidefinite programs in the super-operator form.

\subsection{Parallel approximation of strictly positive semidefinite
  programs}

Parallel algorithms for approximately solving positive linear programs
have been given by Luby and Nisan \cite{LubyN93} and Young
\cite{Young01}.
To our knowledge, an analogous problem for semidefinite programs has
not been considered.

The algorithm from Section~\ref{sec:algorithm} can be used to
approximate, in parallel, strictly positive instances of semidefinite
programs as we now explain.
First, let us note that an arbitrary strictly positive semidefinite
program in the super-operator form can be transformed into one of the
following simpler form:
\begin{center}
  \begin{minipage}{3in}
    \centerline{\underline{Primal}}\vspace{-7mm}
    \begin{align*}
      \text{maximize:}\quad & \tr(Y)\\
      \text{subject to:}\quad & \Phi(Y) \leq \I,\\
      & Y\geq 0.
    \end{align*}
  \end{minipage}
  \begin{minipage}{3in}
    \centerline{\underline{Dual}}\vspace{-7mm}
    \begin{align*}
      \text{minimize:}\quad & \tr(X)\\
      \text{subject to:}\quad & \Phi^{\ast}(X) \geq \I,\\
      & X\geq 0.
    \end{align*}
  \end{minipage}
\end{center}

\noindent
This may be done by defining
\[
\Phi(Y) = A^{-\frac{1}{2}}\,\Psi\left(B^{-\frac{1}{2}} Y
B^{-\frac{1}{2}}\right) A^{-\frac{1}{2}}
\]
for a given problem instance defined by $A>0$, $B>0$, and a strictly
positive super-operator $\Psi$.

Now, to make the connection with the algorithm from the previous
section clear, let us recall that we define $\A=\complex^n$ and
$\B=\complex^m$,
and suppose that the super-operator $\Phi$ that represents the above
semidefinite program takes the form
$\Phi:\lin{\B}\rightarrow\lin{\A}$.
Let us also define $\op{opt}(\Phi)$ to be the optimal value of the
primal problem (which is the same as the optimal value of the dual
problem by strong duality).
It is clear that $\op{opt}(\Phi) > 0$, for some positive scalar
multiple of the identity must be primal feasible.
Let us also recall that we have defined
\[
\alpha(\Phi) 
= \max_{\rho\in\density{\A}} \min_{\sigma\in\density{\B}}
\ip{\rho}{\Phi(\sigma)}.
\]

\begin{prop}
For all strictly positive super-operators $\Phi$ we have
$\alpha(\Phi) = 1/\op{opt}(\Phi)$.
\end{prop}
\begin{proof}
Let $(\rho,\sigma)$ be an equilibrium point of $\Phi$, meaning that
\[
\lambda_1(\Phi(\sigma)) = \ip{\rho}{\Phi(\sigma)} =
\lambda_m(\Phi^{\ast}(\rho)) = \alpha(\Phi).
\]
The assumption that $\Phi$ is strictly positive implies that
$\alpha(\Phi)$ is positive.

We now observe that $\sigma/\alpha(\Phi)$ is primal feasible, as it is
positive semidefinite and satisfies
\[
\lambda_1\left(\Phi\left(\sigma/\alpha(\Phi)\right)\right) = 1,
\]
which implies $\Phi\left(\sigma/\alpha(\Phi)\right) \leq \I_{\A}$.
Likewise, $\rho/\alpha(\Phi)$ is dual feasible as it is positive
semidefinite and satisfies
\[
\lambda_m(\Phi^{\ast}(\rho/\alpha(\Phi))) = 1,
\]
which implies $\Phi^{\ast}(\rho/\alpha(\Phi))\geq \I_{\B}$.
Both result in the same objective value $1/\alpha(\Phi)$, and so the
proposition follows by (weak) duality.
\end{proof}

It follows that the algorithm from Section~\ref{sec:algorithm} may be
used to approximate $\op{opt}(\Phi)$, albeit with limited accuracy for
some choices of $\Phi$, by taking the reciprocal of the value
of the game associated with $\Phi$.
To be more specific, let $R$ be the Choi-Jamio{\l}kowski
representation of the super-operator $\Phi$ as discussed in
Section~\ref{sec:additional-notation}.
Let us write $\widetilde{\alpha}(\Phi)$ to denote the approximate
value of the game described by $R$ that results from the algorithm's
$\varepsilon$-approximate equilibrium point of $R$, and let us also
write $\widetilde{\op{opt}}(\Phi) = 1/\widetilde{\alpha}(\Phi)$.
We then have
\[
\left(1 - \frac{\varepsilon \norm{R}}{\alpha(\Phi)}\right) \alpha(\Phi)
\leq
\widetilde{\alpha}(\Phi)
\leq
\left(1 + \frac{\varepsilon\norm{R}}{\alpha(\Phi)}\right) \alpha(\Phi)
\]
and therefore
\[
\left(1 + \frac{\varepsilon\norm{R}}{\alpha(\Phi)}\right)^{-1}
\op{opt}(\Phi) \leq \widetilde{\op{opt}}(\Phi) \leq
\left(1 - \frac{\varepsilon\norm{R}}{\alpha(\Phi)}\right)^{-1}
\op{opt}(\Phi).
\]

For choices of $\Phi$ for which $\alpha(\Phi)$ is large and $\norm{R}$
is small (bounded below and above by constants, say), a reasonable
approximation to $\op{opt}(\Phi)$ may be obtained.
For many choices of $\Phi$, however, our method is clearly not
suitable, and we believe it is an interesting problem for future
research to find more accurate parallel algorithms for this problem.

\section{Conclusion}

In this paper we have shown that equilibrium points of non-interactive
zero-sum quantum games can be efficiently computed in parallel, using
the multiplicative weights update method.
As a consequence, we have that one-turn quantum refereed games can be
simulated in polynomial space, or
$\class{QRG(1)}\subseteq\class{PSPACE}$.
We have also illustrated the connection between values of quantum
games and positive instances of semidefinite programming problems.

The main open question that we wish to raise concerns the existence of
efficient parallel algorithms for positive instances of semidefinite
programming problems.
The class of such problems for which our algorithm gives accurate
solutions is limited.
To what extent can this task be performed for more general classes?

\subsection*{Acknowledgements}

The parallel algorithm presented in this work was inspired by an
algorithm appearing in an unpublished joint work, concerning classical
games, of Rohit Khandekar and the first author. 
We therefore thank Rohit Khandekar for his implicit and indirect
contribution to this work, and for allowing us to include it in this
paper.
Rahul Jain's research is supported by ARO/NSA USA, and John Watrous's
research is supported by Canada's NSERC and the Canadian Institute for
Advanced Research.

\bibliographystyle{alpha} 


\newcommand{\etalchar}[1]{$^{#1}$}


\end{document}